# Evaluation Pattern on Refugee Crisis

Jiahui Chen, Mengjia Zhou, Bernie Liu

**Summary:** For nowadays severe refugee issue, we primarily set up an evaluation pattern on refugee crisis. Then, we modify our model in the consideration that time and external factors. Besides, we propose a set of policy to facilitate migration movement on optimal route before discussing the extendibility of our model.

**Keywords**：Regression Analysis; Influential Factors; Mini-cost Flows Model; Population Dynamic Theory; Hungarian Algorithm

For evaluation model on refugee crisis, we firstly decide some factors influencing refugee distribution, including comprehensive strength of a country, available resources, and applicants for asylum... Next, we decide the three major factors by applying regression analysis, including refugees in asylum countries, refugees in neighboring countries and available resources, and we get their own weights. Meanwhile, we assemble its calculation to calculating water flowing speed in tube, and get an evaluation model on migration movement speed. After that, we establish a systematic and objective evaluation model.

For the optimal migration movement pattern, we use the mini-cost flows to solve this issue. During this process, we involve the consideration of degree of country ability to accept new refugees and difficulty for refugee reaching different countries, then quantify the data. Finally obtain an optimal migration model. Next, we consider the influence of time and other internal factors on the model. After analysis, environmental factors affected by time are change rate of available resources and that of medical facility, both of which are negatively related to the ability to control. For the method of allocation, we firstly find out the nearby countries and get the difference between change rate of environmental factors and that of refugee amount. Then do the allocation according to the difference by proportion. After using Population Dynamic Model, we conclude the ratio of refugees to citizens in a country will ultimately remain steady. Based on these conclusions, we suggest that if the change rate of environmental factors is bigger than the change rate of refugee amount, the model will be resilient. Then, we also add external factors in model, such as political and assistant factors. To facility refugee movement, we propose a set of refugee policy beneficial to refugee movement by utilizing external power. After that, we strengthen the extendibility of our model, especially to refugee amount. We employ Hungarian Algorithm to set up model on redundant refugee movement, to ensure it can adapt more extreme situation.

# 1 Assumptions of the Model

a) Refugee migration only happens once in a year.
b) The routes between two nodes are straight and there is no barriers on the routes.
c) The statistics we find are reliable.
d) We mainly consider refugees from European countries.
e) Ignoring the number of death during the procedure of movement.
f) When using mini-cost flows method, the minimum cost is the lowest degree of difficulty.

# 2 Symbol description

| Symbol | Definition |
| --- | --- |
| $x_{ij}$ | the $j^{th}$ factor on the $i^{th}$ node |
| $F$ | refugee distribution in exported countries |
| $N_{real}$ | actual amount of refugees |
| $N_{can}$ | Amount of new refugees can be settled down in the countries |
| $T_i$ | the exported amount of refugees in exporting countries |

# 3 Metrics of refugee crises

To decide the influential factors on refugee movement, we firstly let refugees' origin countries, their destinations and countries they pass by be the network nodes abstractly, and the routes they go through are resembled as the lines associating nodes. Therefore, these factors can be subdivided into the influences on nodes and on lines. Countries to be the nodes are as follows: Greece, Bulgaria, Italy, Malta, Cyprus, Spain, Germany, France, Sweden, Ukraine, Switzerland, United Kingdom, Turkey, Egypt, Morocco, Algeria, Libya, Syria and Poland. Details are in Figure 3.1.

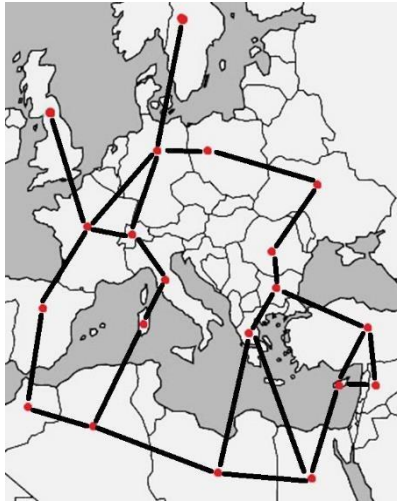

**Figure 3.1 Refugees countries**

Based on our analysis and findings, we primarily determine the potential factors influencing refugee movement as follows: capacity of a country, asylum-seeker, Asylum applications submitted in European countries, available resources (refugees to GDP per capital), amount of refugees in the country and the amount of refugees in neighboring nodes. Meanwhile, the weight among these factors is denoted as $\beta$, then the value of refugees settling on these nodes $y$ is shown in Formula 3.1.

$$y = \beta_0 + \beta_1 x_1 + \cdots + \beta_k x_k \quad \text{(Formula 3.1)}$$

However, the relationship between y and x is multiple linear. So we use the method of multiple linear regression. To obtain the weight of these factors, we firstly standardize the data, the get the relation between the above factors and the amount of refugees on nodes. The statistics of each factor on each node are shown in the below Table 3.1.

Table 3.1 Data of all countries on influence factors

|  | Comprehensive Strength | Applicants for Asylum | Available Resources | Neighboring Refugees | National Refugees |
|---|---|---|---|---|---|
| Greece | 0.369889072 | 42879 | 0.4 | 269180 | 7304 |
| Bulgaria | 0.627270878 | 17895 | 0.62 | 1596860 | 11046 |
| Italy | 1.881850753 | 140629 | 2.62 | 68715 | 93715 |
| Malta | 0.041074615 | 0 | 0.18 | 187843 | 6095 |
| Cyprus | 0.080108169 | 7603 | 0.22 | 1851428 | 5126 |
| Spain | 0.181070464 | 13593 | 0.17 | 253480 | 5798 |
| Germany | 3.950017117 | 455440 | 4.67 | 330625 | 216973 |
| France | 1.475259476 | 309519 | 6.24 | 402625 | 252264 |
| land |  | 226168 | 3.06 | 216973 |  |
| Ukraine | 0.1587 | 867462 | 0.39 | 17923 | 3219 |
| Sweden | 0.405606952 | 83636 | 1.09 | 562952 | 62620 |
| UK | 2.1345 | 154298 | 2.94 | 252264 | 117234 |
| Turkey | 4.404477702 | 1694971 | 81.23 | 16172 | 1587374 |
| Egypt | 0.2764 | 261888 | 22.34 | 40394 | 236090 |
| Morocco | 0.0154 | 4937 | 0.16 | 99926 | 1216 |
| Algeria | 0.0141 | 10381 | 6.68 | 35275 | 94128 |
| Libya | 0.0075 | 400179 | 1.79 | 337522 | 27964 |
| Syria | 0.0145 | 7949617 | 0 | 1592501 | 149140 |
| Poland | 0.276487593 | 29251 | 0.63 | 220192 | 15741 |
| Romania | 0.086784171 | 2874 | 0.11 | 14265 | 2182 |

$x_{ij}$ is denoted as the j$^{th}$ factor on the i$^{th}$ node, while $y_i$ is the value of refugees on the i$^{th}$ node. The process of multiple linear regression includes five steps. Step1：data standardization:

$$\text{let } x'_{ij} = x_{ij} / \max_i \{x_{ij}\} \quad \text{(Formula 3.2)}$$

Step2：build up a linear regression model with k variables, denoted as (Y, Xβ, $δ^2 I_n$). Regression plane equation is given as follows:

$$Y = \begin{bmatrix} y_1 \\ y_2 \\ \vdots \\ y_3 \end{bmatrix}, X = \begin{bmatrix} 1 & x_{11} & x_{12} & \cdots & x_{1k} \\ 1 & x_{21} & x_{22} & \cdots & x_{2k} \\ \vdots & \vdots & \vdots & \ddots & \vdots \\ 1 & x_{n1} & x_{n2} & \cdots & x_{nk} \end{bmatrix}, \beta = \begin{bmatrix} \beta_0 \\ \beta_1 \\ \vdots \\ \beta_k \end{bmatrix}, \varepsilon = \begin{bmatrix} \varepsilon_1 \\ \varepsilon_2 \\ \vdots \\ \varepsilon_n \end{bmatrix}$$

Step3: estimate the value of $\beta_i$ and $\sigma^2$, employ the least squared regression method to get the estimation of $\beta_i \ldots \beta_k$, and then get the sum of squared deviations, as shown in Formula 3.3

$$Q = \sum_{i=1}^{n}(y_i - \beta_0 - \beta_1 x_{i1} - \cdots \beta_k x_{ik})^2 \quad \text{(Formula 3.3)}$$

The result of $\beta_i$ is incorporated into regression model to attain empirical regression model. As shown in Formula 3.4.

$$y = \beta_0 + \beta_1 x_1 + \cdots + \beta_k x_k \quad \text{(Formula 3.4)}$$ With the data
we obtain in programming, the specific equation is shown as bellow.
$$y = 0.0089 + 0.0698 x_1 - 0.0586 x_2 + 1.013 x_3 + 0.128 x_4 - 0.0698 x_5 - 0.6288 x_6$$
Residual plot after regression is in figure 3.2. In the figure, most data is close zero, which meets our expectation.

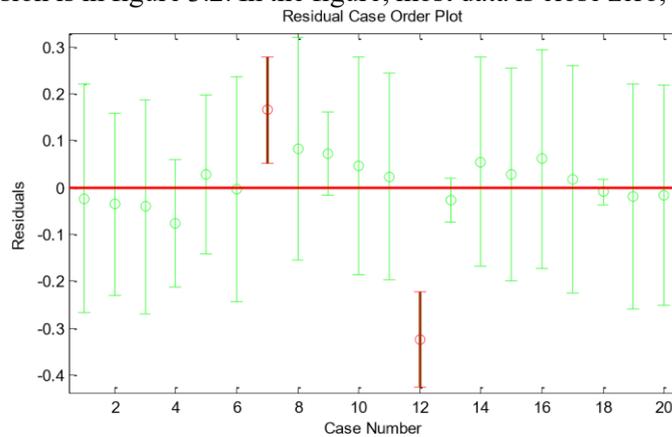

**Figure 3.2 Residual plot after the first regression**

Step4: Apply F test method to test the equation we get, then we reach F=13.5078, which is approved by the test. Then we do t-test on each parameter and reach result in table 3.2. Table 3.2: T-test result on seven variables

| t | constant | $x_2$ | $x_2$ | $x_3$ | $x_4$ | $x_5$ | $x_6$ |
|---|---|---|---|---|---|---|---|
| t-test value | 0.1936 | 0.5068 | -0.3672 | 7.6191 | 0.7788 | -0.6515 | -2.3990 |

In this table, the value of y cannot be obviously explained by some variables, based on the purpose of further stepwise and the difference between coefficient of determination and adjusted coefficient determination is a little bit large. Hence, we consider to modify the aggregation of independent variables and stepwise regression to screen influential factors, as what is shown in step 5.

Step5: Employ stepwise regression, then screen all the factors before get the major influential factors finally.

Stepwise regression is a method of forward selection and backward elimination. For each external variable, as long as it is able to provide significantly explained information, it can enter into the model again, however, for each internal variable, if its F-test is not approved, it can be departed from the model (Wikipedia, 2016). We incorporate data in stepwise regression interactive interface of MATLAB, reaching the result in figure 1.2, from which we can observe that $x_1, x_2, x_5$ are not significantly different, after moving the two variables, the statistical result is given in Figure 3.2.

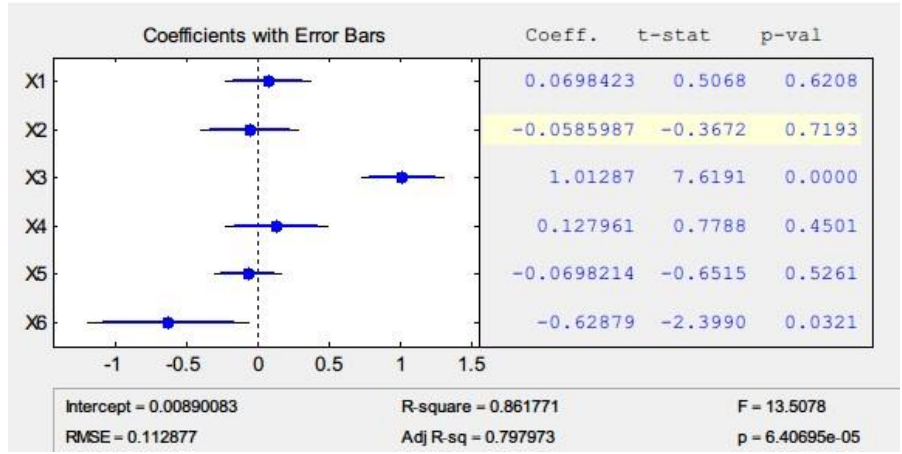

**Figure3.2 T-test result on each variable before being screened out**

Therefore, the major factors influencing refugee distribution on nodes are Asylum applications submitted in European countries, available resources and refugees. The more reliable equation of influential factors and distribution on each node is

$$y = 0.0062 + 0.9936x_3 - 0.0879x_4 - 0.53x_6 \quad \text{(Formula 3.5)}$$

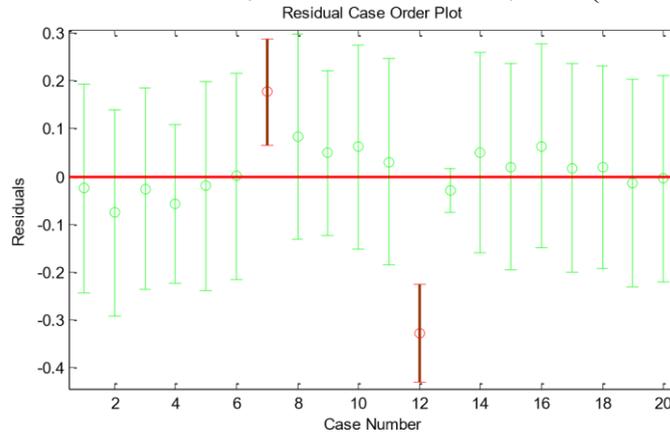

**Figure 3.3 Residual plots after stepwise regression**

After working out the weight of influential factors on nodes, we have to decide the factors on lines, including: migration difficulty, the difference of scores between the neighboring nodes.

We determine the migration difficulty by personal experience. It is obvious to know that the flow rate on the line relates to degree of the difficulty of transportation inversely, while flow rate of line relates to the difference of score on two nodes locating in two sides of a line positively. Applying the relation of the flow rate of tube with pressure and resistance, we finally get the below Formula 3.6:

$$V = \sqrt{\frac{ch}{diff}} \quad \text{(Formula 3.6)}$$

Where diff represents difficulty of transportations among all lines, while ch indicates the difference of scores on two nodes associated by a line. Next, the distribution of nodes we draw in the first step is incorporated in equation seven, and we attain the equation of velocity and each influential factor. Finally, we combine all above formulas, analyze the influential factors generally: factors that affect in refugee movement include difficulty of routes associated with each country, capability of each country, applicants of asylum in each country, available resources in country, number of refugees in neighboring countries and refugees in this country. Besides, the more statistically significant factors are as follows: applicants of asylum in each country, available resources in country and actual number of refugees in this country. Specifically, the more applicants for asylum in this country or the more resources the country can offer, the more refugees will be accepted. Inversely, the less refugees in the country, the worst refugee distribution on this country, hence the less refugees will be accepted. For the flow rates of refugees on all lines, are related to difference of refugee distribution on two nodes of a line and the difficulty of transportation on the route. Specific relation is in formula 3.6, where the larger difference of refugee distribution among nodes, the heavier the flow rate, the more difficult in refugee transportation, and then the lower the speed.

## 4 Flow of refugees

In question two, the optimal refugee movement model we need to figure out is similar to a min-cost flow problem. Primarily, we should decide origin countries of refugee and destination countries, where the conclusion in question one should be employed. Based on the equation we reach in question one, we get the refugee distribution on each country, which reflects the relationship between the amount of submitted applicants in the country and that of actual refugees. It also reflects the connection between available resources in the country and resources have been consumed so far.

Step 1: After using the final result we get in task one, we can work out refugee distribution in each country given in the below table 4.1. In Table 4.1, the odd-numbered lines represent country, while the even-numbered lines indicate quantified value of refugee distribution.

Table 4.1 refugee distribution in each country

| Syria | Egypt | Algeria | Libya | Malta | Morocco |
|---|---|---|---|---|---|
| -0.118 | -0.066 | -0.076 | -0.065 | -0.009 | -0.035 |
| Cyprus | Romania | Ukraine | Spain | Greece | Poland |
| -0.040 | -0.019 | -0.003 | 0.0005 | 0.039 | 0.018 |
| Bulgaria | Switzerland | Italy | Sweden | United Kingdom | France |
| 0.093 | 0 | 0.3487 | 0.227 | 0.397 | 0.194 |
| Germany | Turkey | | | | |
| 0.769 | 0.338 | | | | |

Step 2: Use these quantified value to determine origin country of refugee, destination country and number of refugee. We use ant colony algorithm to classify these countries, the program annex, the results as shown below:

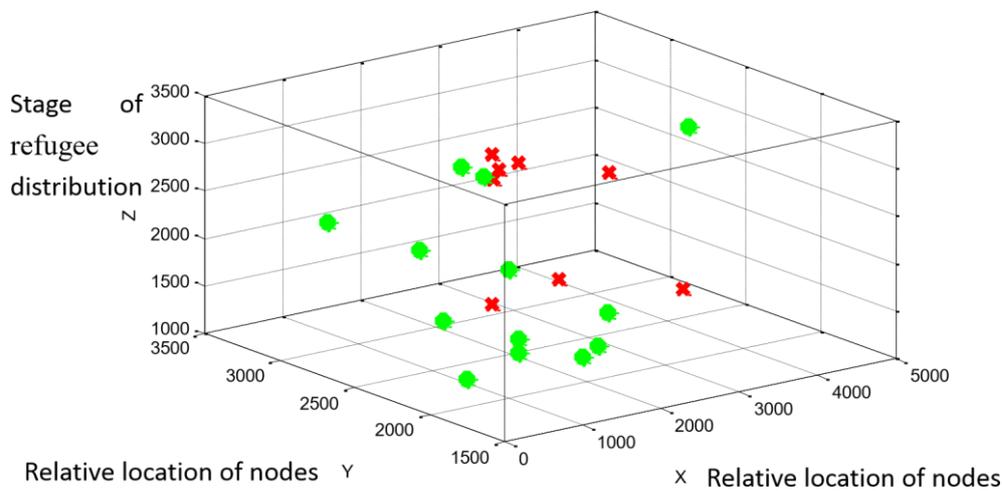

**Figure 4.1 Result of classification between Exporting and importing countries of refugees refugee**

What can b e observed in the first chart is that the exported countries of refugee include Syria, Egypt, Algeria, Libya, Malta, Morocco, Cyprus, Romania and Ukraine. Thereafter, the amount of actual refugee in these countries can be treated as the amount of refugee they produce. Statistics are in Table 4.2.

Table 4.2 The amount of refugee exported from refugee's origin countries
(Popstats.unhcr.org, 2016)

| Country | Syria | Egypt | Algeria | Libya | Malta | Morocco | Cyprus | Romania | Ukraine |
|---|---|---|---|---|---|---|---|---|---|
| Refugee | 149,140 | 236,090 | 94,128 | 27,964 | 6,095 | 1,216 | 5,126 | 2,182 | 150,000 |

(Because of Ukraine War, numerous refugee in Ukraine flee away, therefore the recorded figure of refugee apply for asylum cannot correctly reflect the exported amount of refugee, by searching for the related news

(BBC News, 2016), we decide the number is 150,000）

Spain, Greece, Poland, Bulgaria, Switzerland, Italy, Sweden, United Kingdom, France, Germany and Turkey are countries receiving refugee. We definite the refugee amount that a country is able to accept is

$$N_{can} = N_{real}(1 - f) - N_{real} \qquad \text{(Formula 4.1)}$$

Where f is denoted as refugee distribution in exported countries, $N_{real}$ indicates the actual amount of refugees, and $N_{can}$ represents how many new refugees can be settled down in the countries. Incorporate data of each country into the formula, and then we get the number of new refugee will be accepted by each country in the following Table 4.3.

Table 4.3 estimation on amount of refugees accepted in asylum countries

| Country | Spain | Greece | Poland | Bulgaria | Switzerland | Italy |
|---|---|---|---|---|---|---|
| Accepted refugee | 3 | 297 | 281 | 1,138 | 2 | 50,173 |
| Country | Germany | Turkey | Sweden | United Kingdom | France | |
| Accepted refugee | 723,482 | 810,109 | 42,102 | 77,108 | 60,682 | |

Step3：By using mini-cost flow proposal, we do a proper allocation on refugee settlement in accordance with difficulty on each route, amount of exported refugees on exporting countries and the amount of refugee will be accepted by destination countries to reach the optimal refugee movement model. The six routes given in the task would be an example. As the below figure 1.1 and 1.2 show.

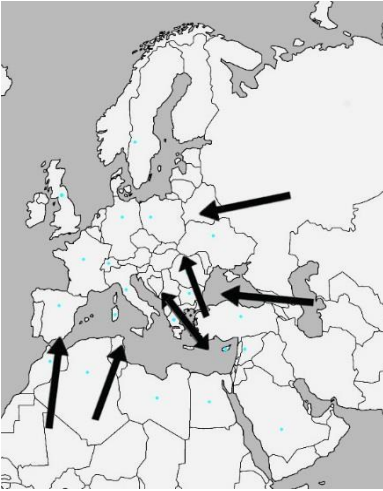
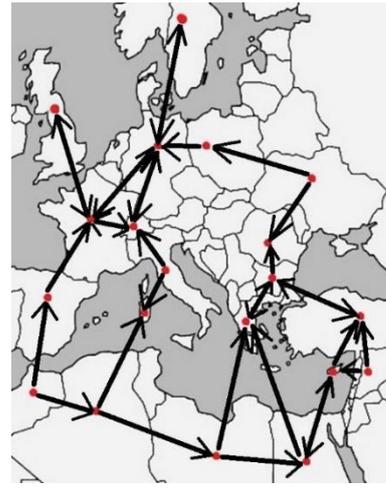

**Figure 4.1 The six major migration routes**　　　　　　**Figure 4.2 The main detail routes**
(Image: Migrants detected entering the EU illegally, Jan-Nov 2015)
Figure 1.2 oriented graph on migration routes The six routes can be
mainly described as follows:

① Western Mediterranean：Morocco→Spain→France→Germany and other countries in western Europe
② Central Mediterranean：Libya→Malta→Italy→France, Germany and other countries in western Europe
③ Ukraine→Poland→Germany
④ Syria→Turkey→Bulgaria→Greece and other countries in eastern Europe
⑤ Egypt→Cyprus→Turkey→Greece and other countries in eastern Europe
⑥ Greece→Bulgaria→countries in eastern Europe

The problem we need to figure out is about min-cost flow. Firstly, we give the weight on each line among nodes. The associated lines in graph have direction, so the graph is a directed graph. We define the calculation method of weights on cost among the associated lines as follows:

$$Df_{tr} = \frac{T_i}{R_{i+1}} \qquad \text{(Formula 4.2)}$$

$$Df_{tt} = \frac{T_{i+1}}{T_i} \quad \text{(Formula 4.3)}$$

$$Df_{rr} = \frac{R_i}{R_{i+1}} \quad \text{(Formula 4.4)}$$

In formula (1), $Df_{tr}$ is denoted as the difficulty on refugee movement between exporting countries and receiving countries, while $T_i$ indicates the exported amount of refugees in exporting countries, and $R_{i+1}$ means the number of refugee can be accepted by next node country. Besides, the ratio of $Df_{tr}$ to $R_{i+1}$ can reflect migration difficulty. In formula (2), $Df_{tt}$ indicates the degree of difficult on migration movement from exporting countries, whereas $T_{i+1}$ reflects the acceptable amount of refugee in next node and $T_i$ is number of refugee that is capable to be accepted by current node. In formula (3), $Df_{rr}$ indicates difficulty on refugee movement among accepting countries, $R_i$ represents the accepted refugee amount on current node, and $R_{i+1}$ suggests the amount of refugee can be accepted by the next node. The ratio of the current node to the next node reflects the influence on migration priority imposed by amount of accepted refugee in next node country and number of refugee will be accepted or exported in current node country, which can assist the minimum path method we are going to use. Even the less the difficulty weight which is similar to the cost in mini-cost flow issue, the larger the priority degree of the route. In addition, capacity on each road ought to be considered. In other words, when thinking about the route to do the easiest movement, we should also decide how to increase flows on each road. Therefore, we assign the number of refugee can be received in next node to be origin capacity of the road. By applying Floy algorithm, we get the shortest path between each two nodes, that is the most reachable route. Then by incorporating capacity matrix, we modify flows constantly and get the matrix matched to the optimal migration graph. The maximum flow in any capacity network is the only one and certain, but its maximum flow f is more than one. Since we have more than one maximum flows f, we have to realize that each arc not only has capacity limitation but also confined cost r---the parameter for cost per unit on each arc. Provided that the maximum flow exists, there occurs a problem for choosing the minimum cost, which is the mini-cost flows. The way to find the maximum flow is start from an accessible flow, find an augmenting path P of this flow, and modify f when following P. For new accessible flows, we try to find its augmenting road, doing the circulation until augmenting paths do not exist. Next, we work out the mini-cost flows issue.

If f is the minimum cost among f1 accessible flows, while p is the least costed augmenting path among all augmenting paths. Then we follow p to adjust f, finally get the accessible flow f, which is the minimum cost among all f1 accessible paths. Therefore, when f is the maximum flow, it is the mini-cost flows we need to figure out.

Calculating processes are as follows:

1. For net G=[V,E,C,W], we give initial flow whose value is zero.

2. Formulate the increasing-flow network G′=[V′, E′, W′] for this flow. The apex on G′ is the same as that of G: V′=V. If f(u,v)<c(u,v) in, then set a side (u,v) around G, w(u,v)=w(u,v). If in G, f(u,v)>0, then set a side (v,u) in G′, w′(v,u)=-w(u,v).

3. If G′ does not exist on the path from x to y, then the flow of G will be the mini-cost flow, and stop calculation. Otherwise, we use labelling method to find the shortest path from x to y.

4. Based on P, we add flows on G: for each side (u,v) on P, if G is in (u,v), then (u,v) will increase flows; while G is in (v,u), (v,u) will decrease flows. After increasing or decreasing flows, any side with c(e) ≥ f(e) ≥ 0 should be ensured.

5. According to the value L(v) signed on each apex on the shortest path, we modify weights of all sides around G as the following instruction:

$$L(u)-L(v)+w(e) \rightarrow w(e) \quad \text{(Formula 1.1)}$$

6. Treat new flows as initial flows, turn to step 2. We take Turkey, Syria, Egypt, Cyprus and Libya as examples, calculate the best migration route from Libya to Turkey. The initial difficulty matrix matched to the graph is:

$$w = \begin{bmatrix} 0 & 4 & 1 & M & M \\ M & 0 & M & 6 & 1 \\ M & 2 & 0 & 3 & M \\ M & M & M & 0 & 2 \\ M & M & M & M & 0 \end{bmatrix}$$

The origin capacity matrix in accordance with graph is

$$x = \begin{bmatrix} M & 10 & 8 & 0 & 0 \\ 0 & M & 0 & 2 & 7 \\ 0 & 5 & M & 10 & 0 \\ 0 & 0 & 0 & M & 4 \\ 0 & 0 & 0 & 0 & M \end{bmatrix}$$

By using the mini-cost flows model, we get the optimal migration movement matrix as follows: Through

$$f = \begin{bmatrix} 0 & 6 & 16 & 0 & 0 \\ 0 & 0 & 0 & 0 & 14 \\ 0 & 8 & 0 & 8 & 0 \\ 0 & 0 & 0 & 0 & 8 \\ 0 & 0 & 0 & 0 & 0 \end{bmatrix}$$

Subsequently, we have to determine the refugee assignation on travel route from origin countries, since some origin countries possess several migration routes. Take an origin country node as an example, we set up the migration routes model as below.

Assume a origin country node of refugee is x, the reachable destination node is $Y_i$, and the degree of priority for $X \to Y_i$ is $Q_i$. Then the allocated people on route $X \to Y_i$ is:

$$P_i = \frac{Q_i}{\sum_{i=1}^{k} Q_i} \quad \text{(Formula 4.5)}$$

Input the date into the mentioned mode, we might draw the best migration movement scheme.

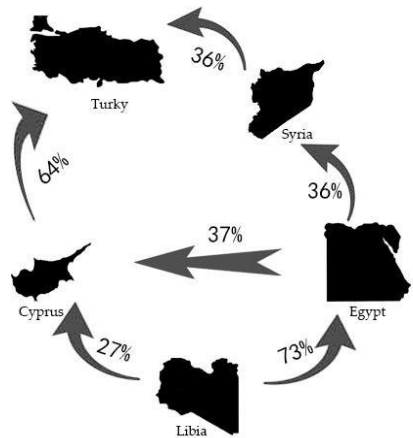

Finally, all the countries of the nodes in the model data into them, can get all the bestmigration routes, as shown in Figure 4.4.

**Figure4.3 The optimal migration routes**

Figure 4.4 The optimal routes to all the countries

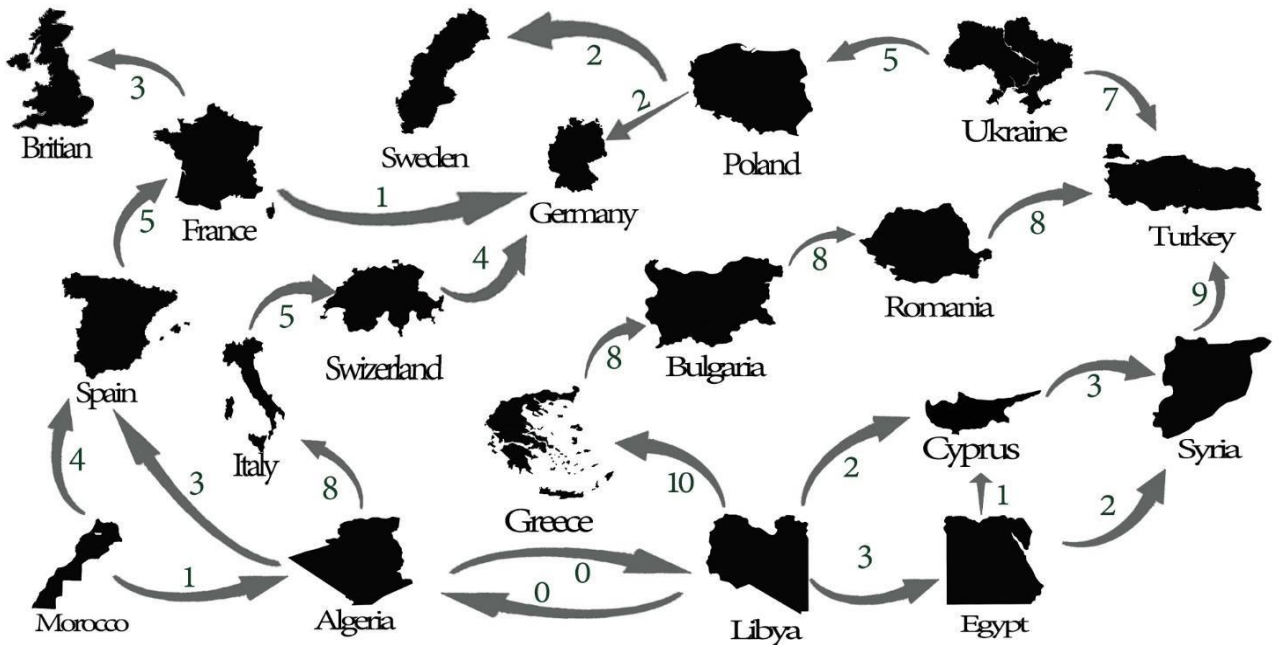

# 5 Dynamics of the crisis

In question three, we firstly need to decide environmental factors influencing our scheme. The major environmental factors include available resources (water, bed, clothes, shoes…) and medical facilities which are also the supplies need to be prepositioned and allocated for priority. Then we elaborate how environmental factors affect our movement scheme as time goes by. We choose Spain to be our researched subject. Secondly, we decide how the supplies be allocated when resources in Spain is shortage. Then, we employ population movement model to check whether the ratio of refugees to citizens will be stable ultimately, and finally protect ability to control resources from time effects.

Step 1: find the relationship between ability to control and environmental factors.

Table 3.1 Relationship between ability to control and environmental factors

| Year | 2010 | 2011 | 2012 | 2013 | 2014 |
|---|---|---|---|---|---|
| Rate of change in medical facility | 12% | 14% | 18% | 21% | 26% |
| Available resource change rate | 11% | 16% | 17% | 24% | 25% |
| Ability to control | 9 | 7 | 8 | 3 | 1 |

Inputting the above data into Linear Regression Algorithm, we get figures and formulas as below:

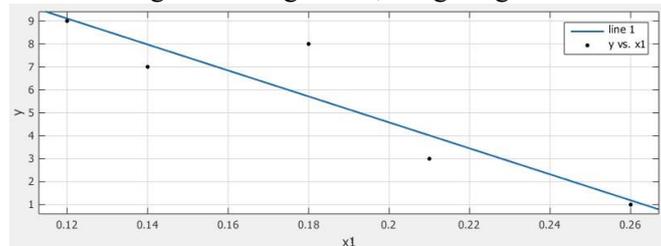

**Figure 3.1 Relationship between ability to control and rate of change in medical facility**

The equation of line in figure 3.1 is y=-56.57*x1+15.9

Where y is the ability to control and x1 is the rate of change in medical facility.

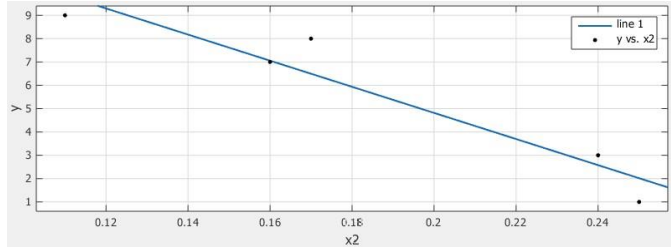

**Figure 3.2 Relationship between ability to control and available resource change rate**

The equation of line in figure 3.2 is

$$Y = -55.98 \cdot x_2 + 16.01 \quad \text{(Formula 1.1)}$$

Where Y is the ability to control, x2 is available resource change rate. From the above analysis, we can figure out that the ability to control is negatively related to environmental factors.

Step 2: Next, we decide how to preposition material resources and how to allocate them.

From our finding results (details are in appendix 3.1), we know the annual change rate of refugee in 2014 is 46.94%, while its rate of change in medical facility is 24% and available resource change rate is 25%. Obviously, the change rate of refugee is about 22% higher than change rates of the both environmental factors. Hence, Spain have to import resource from neighboring countries. Non-governmental Organization (NGO) has the responsibility to grant national resource shortage issue (Wikipedia, 2016). Hence, under the intervention of NGO, Spain is able to attain resources for nearby countries without any barrier, while the nearby countries are Portugal Italy and France.

Table 3.2 Statistic of Environmental factors in neighboring countries

| Country | Portugal | Italy | France |
|---|---|---|---|
| Change rate of refugee | -22.11% | -54.99% | 8.43% |
| Available resource change rate | 6% | 8% | 11% |
| Change rate of medical facility | 7% | 8.5% | 11.7% |

In the table, the last two change rates in Portugal, Italy and France are larger than their first one change rate, hence, Spain can get resources from these three countries. Since the differences between change rate of refugee and other two change rates in three countries are 30%, 60% and 10% respectively. Therefore, Portugal, Italy and France can transport supplies in the proportion of 6:3:1 to Spain.

Step 3: It is known to all that proportion of population migration and emigration is fixed as time goes by, then refugees entering can be considered as population migration. The procedure of population movement is given below, with Canada being an example.

It can be described by matrix multiplication.

$$x_s = \begin{bmatrix} x_{as} \\ x_{bs} \end{bmatrix}$$

Where $x_a$ is the proportion of citizens in a country, $x_b$ is the is the proportion of refugees, s is the sequence of years. The initial status when k=0 is

$$x_0 = \begin{bmatrix} x_{a0} \\ x_{b0} \end{bmatrix} = \begin{bmatrix} 0.7 \\ 0.3 \end{bmatrix}$$

After a year, the amount of citizens is

$$x_{a1} = (1 - 0.05)x_{a0} + 0.01 x_{b0}$$

The amount of refugees is:

$$x_{b1} = 0.05 x_{a0} + (1 - 0.01) x_{b0}$$

The above formula can be written in matrix multiplication as below:

$$x_1 = \begin{bmatrix} x_{a1} \\ x_{b1} \end{bmatrix} = \begin{bmatrix} 0.95 & 0.01 \\ 0.05 & 0.99 \end{bmatrix} \cdot \begin{bmatrix} 0.3 \\ 0.7 \end{bmatrix} = A x_0 = \begin{bmatrix} 0.29 \\ 0.71 \end{bmatrix}$$

From initial time to year k, the relation remains the same, then the above equation can be extended as follows:
Extend time k infinitely, the proportion will remain the same as shown in Figure 5.3.

As time goes by, the ratio of citizens to refugees in any country will be stable, which can be seen in the following three figures. (Matlab code and referenced data are in appendix 5.2)

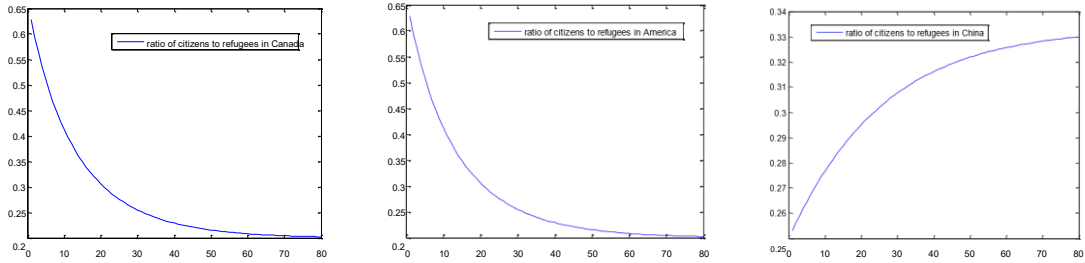

**Figure 5.3 Variety of Canada   Figure 5.4 Variety of America   Figure 5.5 Variety of China**

Since ratio of citizens to refugees in any country will remain stable finally, if we guarantee the resource change rate in the country is higher than its refugees change rate after accepting help from other countries, the ability to control in this nation will be ensured.

## 6 Policy to support refugee model

Our team simplify the six routes from the Middle East to West Mediterranean, Central Mediterranean, Eastern Mediterranean, West Balkans, Eastern Borders, and Albania to Greece as nodes and lines in network. Next, we select the three major influential factors to build up a linear regression equation:

$$y = 0.0089 + 0.0698x_1 - 0.0586x_2 + 1.013x_3 + 0.128x_4 - 0.0698x_5 - 0.6288x_6$$

According the difficulty on each route, the amount of exported refugees from exporting countries and the amount of refugees can be accepted by asylum countries, we allocate refugees' settlement properly by using mini-cost flows. Then we get the optimal migration model as Figure 1.1, based on which we give out the follow policies:

1. After screening out less influential factors on migration, we reach major factors including applicants for asylum, available resources in a country and refugees in the country. Therefore, we suggest asylum countries in Europe such as Spain, Greece, Bulgaria, Switzerland, Italy, Sweden, United Kingdom, France, Germany and Turkey approve more applicants when each individual has surplus resources independently.

2. Through observing the model, we notice that Syria is the major origin country of refugees. Thereafter we recommend non-commercial organization like Red Cross offer humanity help to refugees in Syria or set up assistant houses on their way to Turkey to decrease the degree of danger.

3. Since the increase of refugee influx, cost of commute will increase and environment will be deteriorated. Hence, we need to do various degree of urban planning and design on countries accepting refugees. Based on the rank of overall capability for a country and double factors analysis on capacity of refugee, we reach the amount of refugees needed to be dealt with in Cyprus, to ensure that it reaches the optimal capacity and maximum capacity. After selecting and screening out data, we find that the number of refugees in Turkey is the largest but its overall capacity is not strong enough to burden so many refugees. Therefore, we suppose assign some refugees in its neighboring countries, or control the amount of entering refugees.

4. According to the model we have and happiness model proposed by OLS in *City Scale, Sense of Happiness and optimization of immigration space*, personality and occupation of refugee may affect Gross National Happiness. Hence we recommend increase job occupation for refugees from Turkey and Cyrus, as well as increase houses for submitted applicants which then increase its flow.。

5. Considering the health and safety of refugees and local citizens, we suppose increase assistant houses, set up safe entry point in the optimal route through these countries: Turkey, Cyprus, Egypt, Libya and Syria, and provide disinfection and disease therapy for refugees.

6. Implement immigration political participation policy, give refugees the right to voice their ideas in policy, which will reduce refugees' discrimination and then decrease national conflicts, furtherly decrease the number of refugees in the future.

## 7 Exogenous events

The occurrence of external event will impose great effects on each influential factor for each country, furthermore influencing the weight on each factor matched to distribution of refugee on nodes. The selected factors include national comprehensive strength, number of applicants for asylum, available resources, refugee amount on neighboring nodes and actual amount of refugees in the country. When external aspects influence these influential factors and change statistic obviously, the ability to explain refugee distribution on nodes for factors, which influence the cost matrix and capacity matrix in mini-cost model. The changing value of each element in these two matrixes leads to the alteration of migration movement scheme. Thereafter, the changing parameter is mainly element value in cost matrix and capacity matrix, that is the capacity and difficulty of each route.

After analyzing the theory, we study the case in France and Belgium.

Case 1: Input the amount of refugees who influx in France, the influx of refugees before and after the attack is as follows:

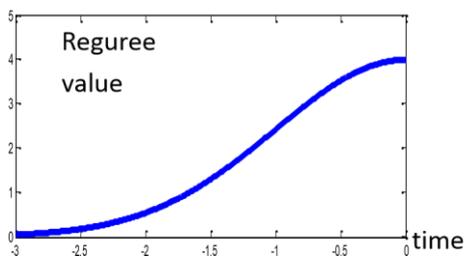 Paris Terrestrial attack → 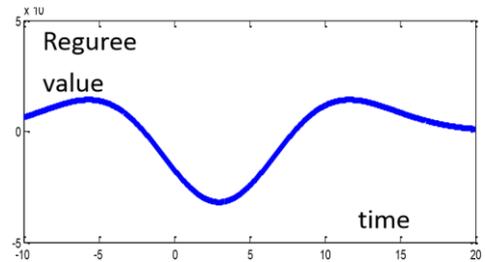

**Figure 7.1 Change of French Refugees before Attack**  **Figure 7.2 Change of French Refugees after Attack**

Table 7.1 The Parameter Reflection to Parameters of Paris Attacks and Result Alternation

|  | Before Attack | After Attack |
|---|---|---|
|  | $\begin{bmatrix} M & M & M & M & M \\ 5 & M & M & M & M \\ M & 5 & M & M & M \\ M & M & 1 & M & M \\ M & M & 5 & 7 & M \end{bmatrix}$ | $\begin{bmatrix} M & M & M & M & M \\ 5 & M & M & M & M \\ M & 5 & M & M & M \\ M & M & 6 & M & M \\ M & M & 5 & 7 & M \end{bmatrix}$ |
| Optimal route | 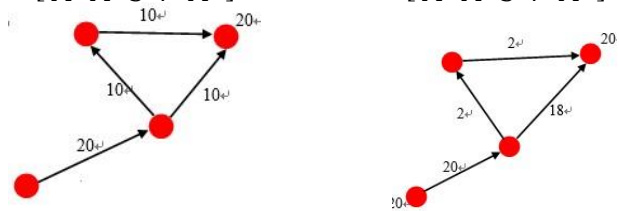 |  |

Capacity Matrix

In graph 7.1 and 7.2, the left one assumes the influx trend of Paris refugee before Paris Attack. Meanwhile, the right graph indicates the influx trend after Paris Attack. It is apparent that the amount declines dramatically after the attack. However, after a while of time, it rebounds to the previous amount.

Case 2: In Belgium, there are two external events affecting refugee amount.
① Refugee event happened nearby Belgium.
② The announcement on relative policy in Belgium imposes great limitation on receiving amount of refugee.
③ After Paris Attack, Belgium is lockdown.

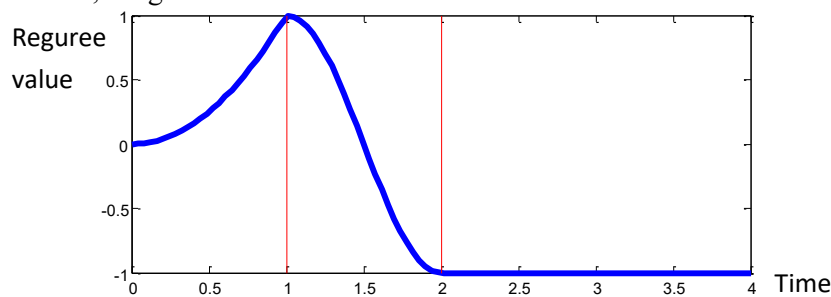

**Figure 7.3 Refugees Population Movement Trend on Belgium**

As shown in above figure, stage 0-1 indicates the change of refugee amount in Belgium after refugee event happening. While stage 1-2 records the change of refugee amount after announcement on relative policy in Belgium. Stage 2-4 illustrates the change of refugee amount after lockdown in Belgium. In stage 0-1, the major change of parameter is cost matrix. In stage 1-2, the major change of parameter is capacity matrix. When X=2, the alternation parameter is capacity matrix.

The selection of initial factors we get in task one on refugee in neighboring countries, including refugee amount in neighboring countries, is the same as the initial model we get in task 1:

$$y = 0.0089 + 0.0698x_1 - 0.0586x_2 + 1.013x_3 + 0.128x_4 - 0.0698x_5 - 0.6288x_6$$

From the formula, we can observe that refugee amount on neighboring countries affects that on node countries. The increase in refugee amount on neighboring countries leads to the increase in that on node countries, which decreases the travelling cost on this route in cost matrix from mini-flows cost model, furtherly affecting the optimal migration model.

3、Through adding other assistant factors, we transform the interference created by external aspects. For instance, the occurrence of Paris Attack increases the traveling cost on the way to this country in cost matrix. According to our policy, however, some Non-Governmental Organizations and assistant institutions will help refugees during their movement. So, we consider Non-Governmental Organization as a node.

When calculating difficulty of route, we insert external factors, and deal with the interference from external factors, as what the below formula shows.

$$\text{Diff} = \text{Out} \times (T_i - R_{i+1}) \quad \text{(Formula 1.1)}$$

Where Diff indicates route difficulty, $T_i$ represents current amount of refugees on nodes and $R_{i+1}$ is amount of refugees on the next node.

We let Egypt, Cyprus, Turkey, Syria and Libya be the examples. Before adding externally intervened factors, the difficulty matrix and optimal route model is:

Table 7.2 Indication of model affected by externally assistant factors

| | Before Adding External Help | After Adding External Help |
|---|---|---|
| Difficulty Matrix | $w_1 = \begin{bmatrix} 0 & 4 & 1 & M & M \\ M & 0 & M & 6 & 1 \\ M & 2 & 0 & 7 & M \\ M & M & M & 0 & 2 \\ M & M & M & M & 0 \end{bmatrix}$ | $w_2 = \begin{bmatrix} 0 & 4 & 1 & M & M \\ M & 0 & M & 6 & 1 \\ M & 2 & 0 & 3 & M \\ M & M & M & 0 & 2 \\ M & M & M & M & 0 \end{bmatrix}$ |
| Optimal Route | 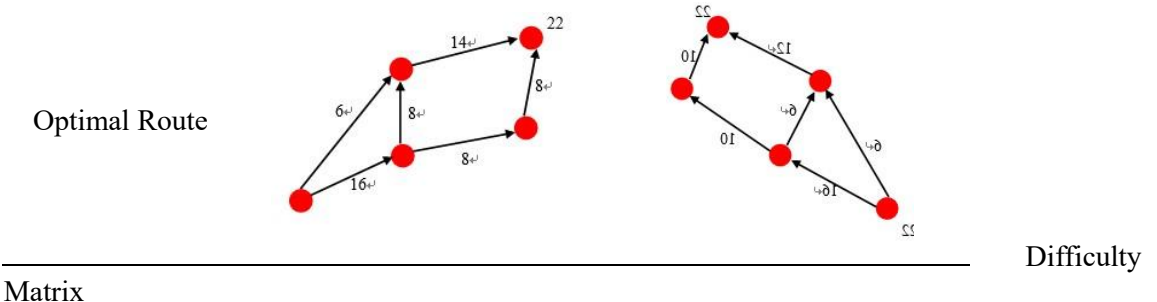 | |

From above tables, the migration model with external factors is more practical.

Adding new nodes in the model indicates the positive influence for Non-Governmental Organization. For instance, on roads between Syria and Turkey, Non-Governmental Organization assists refugee migration. After adding new nodes, the changes of difficulty matrix and capacity matrix are in below table, while the alteration of optimal route is in the last row of Table 1.1.

Table 7.3 Model's reflection to non-governmental organization（by adding nodes）

| Before Adding Nodes | After Adding Nodes |
|---|---|
| $\begin{bmatrix} M & 0 & 0 & 0 \\ 6 & M & 0 & 0 \\ 4 & 2 & M & 0 \\ 0 & 7 & 6 & M \end{bmatrix}$ 13 | $\begin{bmatrix} M & 0 & 0 & 0 & 0 \\ 8 & M & 0 & 0 & 0 \\ 2 & 2 & M & 0 & 0 \\ 0 & 6 & 0 & M & 0 \\ 0 & 0 & 4 & 6 & M \end{bmatrix}$ |

|  | Capacity Matrix | |
|---|---|---|
| Matrix | $\begin{bmatrix} 0 & M & M & M \\ 4 & 0 & M & M \\ 6 & 8 & 0 & M \\ 0 & 3 & 4 & 0 \end{bmatrix}$ | $\begin{bmatrix} 0 & M & M & M \\ 2 & 0 & M & M \\ 8 & 8 & 0 & M \\ M & M & 6 & 4 & 0 \end{bmatrix}$ Difficulty |
| Optimal Route | (graph with nodes, edges labeled 4, 4, 10, 1, 6, 5, 10) | (graph with nodes, edges labeled 2, 4, 10, 2, 10, 8, 6, 6, 10) |

## 8 Scalability

For the extendibility of our model, we firstly consider the extension of model in different regions, different refugee amount and different external factors. We analyze extendibility of our model in these factors as below.

1. With regional expansion, the change of accurate rate for model modifies the amount of refugees. However, it doesn't matter on calculation for optimal paths. Besides changing nodes number of model, the alteration of data from subjective districts will also lead to the change of influential factors, however, these can be dealt with by the model and not affect final result.

2. For the increasing amount of refugees, we assume all refugees can move to some districts. Hence, when refugee is overcrowding and applicants' surplus the capacity amount in asylum countries, our model becomes useless. To enhance extendibility of our model on refugee amount, we propose ways to split flows for redundant population. Different countries have various attitude on receiving redundant migration. Thereafter, we need to involve attitudes towards countries as well as its capacity to receiving redundant refugees and various amount of refugees accepted by different countries. We transform it to a assignment problem, reach the result by applying Hungarian Algorithm.

Step1: As the below table shows, provided that there are three countries (Libya, Algeria and Syria) with unsettled refugees, while another three countries (Germany, France and England) are able to accept new refugees but they nearly reach their own maximum capacity. Out of humanitarianism, they will receive new refugees until reaching the limitation. The number of refugees they can still accept is as follows. Table 6.1 Number of Refugees Can by Accepted（unit: 10,000）(BBC, 2016)

|  | United Kingdom | France | Germany |
|---|---|---|---|
| Libya | 1 | 4 | 2 |
| Algeria | 3 | 2 | 2 |
| Syria | 1 | 1 | 5 |

Step2: Transform above table to matrix can be used by Hungarian Algorithm. Hungarian Method in Assignment Problem (Wikipedia 2016). The elements in the rows of elements respectively subtracting the minimum value of the Bank. And then, on the basis of the elements in each column and then subtracting the minimum value of the column.

$$\begin{bmatrix} 1 & 4 & 2 \\ 3 & 2 & 2 \\ 1 & 1 & 5 \end{bmatrix} \rightarrow \begin{bmatrix} 0 & 3 & 1 \\ 1 & 0 & 0 \\ 0 & 0 & 4 \end{bmatrix}$$

Identify independent zero elements, and mark.

Determine whether there is a progressive independent lines containing 0 elements firstly. If any, continue to press the line processing. If not, will have to determine whether there is a column-by-column contains 0 independent elements. If so, press the column processing continues. If neither independent line contains 0 elements, there is no independent columns containing 0 elements, the process continues to press the line. And only [0 3 1] as an independent line contains 0 elements in the resulting matrix. When press line processing, if a row with separate elements 0, 0 to the element marked as a, the column where the remaining 0 element is marked as b. Otherwise, the Bank temporarily crossed the deal behind OK. 0 contains all the elements of an independent line after processing, then come back and deal with two containing two or more elements in row 0: 0 and optionally do a tag, and then the remaining 0 row and column of elements the remaining elements are marked b. When handling columns, if a column has a separate element, the element that is marked as a, 0 where the rows of the remaining element is marked as 0 b; otherwise, this column is temporarily over the deal behind columns. The column contains all the elements of independent 0 after processing. Then come back and deal with two elements of two or more columns containing 0: 0 and optionally do a mark. Then the 0 column of the remaining elements and the 0 row of the remaining elements are marked b. (4) Repeat the process to obtain independent zero elements (labeled "0" of a). The calculation procedure is as follows.

$$\begin{bmatrix} 0 & 3 & 1 \\ 1 & 0 & 0 \\ 0 & 0 & 4 \end{bmatrix} \longrightarrow \begin{bmatrix} 0_a & 3 & 1 \\ 1 & 0_a & 0_b \\ 0_b & 0_b & 0 \end{bmatrix}$$

$$\longrightarrow \begin{bmatrix} 0_a & 3 & 1 \\ 1 & 0_a & 0_b \\ 0_b & 0_b & 0 \end{bmatrix} \longrightarrow \begin{bmatrix} 0_a & 3 & 1 \\ 1 & 0_a & 0_b \\ 0_b & 0_b & 0_a \end{bmatrix}$$

Step 3: If the number of independent zero elements equals to matrix order, we get the optimal solution. However, if it is smaller than matrix order, we continue following steps: (1) Mark c for the lines that do not have a mark of a. (2) For the columns that have been marked b, the column marked where the mark c. (3)Having been marked in the column c, the mark of a line where the mark c. (3)Do not marked c scribe line on a column marked c scribe. (4) Find an element $x_{min}$ with the minimum value, each element on rows and lines uncovered by the line should be subtracted by it. (In uncovered area, if the order of row is smaller than the order of column, then subtracted by row, otherwise subtracted by column.

Step 4: Since negative elements will occur, we should add this element $x_{min}$ with the minimum value on the rows or lines that have negative elements to eliminate them. Then back to step 2，decide the amount of independent of zero elements. Repeat the above steps until find out the optimal solution.

As the number independent zero elements is equal to the amount of matrix order 3, so the optimal solution is:

$$\begin{bmatrix} 0 & 3 & 1 \\ 1 & 0 & 0 \\ 0 & 0 & 0 \end{bmatrix}$$

So the best solution is that Libyan refugees to go to England, refugees in Algeria to France, Syrian refugees to Germany. Since Hungarian Algorithm can decide the minimum value of consumption. Hence we select numbers which are inverse to data, and get a combination with the maximum population. Hence, we get the optimal allocation for refugee. Calculation procedures are as follows.

3. For influence on extendibility in model from outer factors, we carefully evaluate its effect, then we add political factors and assistant factors into the model. Therefore, when new external elements occur, we can extend our model by increasing outer factors.

# 9 Evaluation of Model

Strengths:

1、 Based on analyze problems accurately and clearly, we set up a proper and scientific model on calculating the optimal path, and provide keys for optimal migration issue.

2、According to assumption, we set up regression model, decide amount of applicants for asylum, available resource in a country and its actual refugee to be the three major influential factors.

3、We promote and apply mini-cost flows model, and program it in MATLAB, which is full of practical value.
4、We add priority allocating resource and medical facility into consideration, making the model more practical.

Weakness:

1、The routes we consider and the data we collect are not enough.

2、Refugee movement in reality is different with that in our model. So in our calculation, we use the predicted value as approximate value in calculation, which differs from actual value to some extent.

3、The introduction variables are too many when building up a model, which leads to Curse of Dimensionality easily and difficult to proceed programming.

Expansion of the model:

It is easy to find that this question is about linear programming. We build up linear programming model to determine influential factors, select the optimal route by using mini-cost flows and adding of factor optimization model by applying population migration model. After analyze the model carefully we find out that it is not only applicable on refugee migration problem but also influence solving linear, non-linear or dynamic programming issues effectively. Programming is an important branch of operational research. It plays an important role in working out industrial production, economic plan and management on man-machine system. If decision makers want to get a well-rounded optimal result, they must abstract concept, analyze connection and put each type of factors into programming model. The model we reach is a typical model about programming problem with various application, such as the way to allocate limited capital in various financial instruments, how to position an industry when thinking its distance to raw material zone and service zone, and so on. Overall, programming model have various functions in industry, business, transportation, engineer and politic management.

# 10 Policy recommendation letter

Dear UN Secretary General and the Chie of Migration,

Refugee issue threats world peace and mutual development. The unique challenges these crisis possess are supposed to be dealt with seriously by effective policies. Therefore, our team do modeling analysis on the crisis, and educe the suggestions as follows:

1. Strengthen reginal facility construction. After selection, we reach the major influential factors including number of applicants in asylum countries, available resources in the country and its actual amount of refugee. The national comprehensive strength in Turkey enables it to accommodate more refugees. Hence, to accept more refugees, more entry points and constructional facilities need to be built in Turkey.

2. Enhance safety on migration routes. Syria is the major origin country of refugees with 149,140 being the figure. Thereafter, we suggest government ought to unite with non-commercial organization such as the Red Cross for assisting refugees in Syria humanitarianly and reducing degree of difficulty on migration by instructing refugee flows move to Eastern Mediterranean.

3. Control the flows of refugee. Our model indicates that Cyprus needs proper management to reach the optimal scale or the maximum scale in cities. From data we collect and select, we discover that Turkey accepts the largest amount of refuge but its comprehensive strength is not strong enough to burden so many refugees. Hence, we recommend government allocate part of the refugees in neighboring countries or control the entering refugee amount or flows.

4. Create new job occupation for refugees. We advise Turkey and Cyprus offer specific occupation for refugees. Besides, we suggest they increase houses to hike the number of applicants, which then increases flows of refugees. In addition, more schools should be established to provide educational resources for refugees.

5. Facilitate movement of refugees on the optimal routes. Set safety stations on entrance and exit of each passing through countries, as well as add assistant sites on roads out of considering the health and safety of refugees and local residents.

6. Implement refugee political participation policy. Then local conflicts created by refugee discrimination will be reduced, which furtherly decrease number of refugees in the future to work out refugee issue better.

<div align="right">Yours Sincerely,<br>ICM-RUN</div>